\documentstyle{article}

         \def\ba{\begin{array}}
         \def\ea{\end{array}}
         \def\be{\begin{equation}}
         \def\ee{\end{equation}}
         \def\det{{\rm det}}
         \def\GL{{\rm GL}}
         \def\SL{{\rm SL}}
         
         \def\sl{{\rm sl}}
         \def\half{{1 \over 2}}
         \def\IGL{{\rm IGL}}
         
         \def\Uq#1{{\rm U}_q \left( #1 \right) }
         \def\Uh#1{{\rm U}_h \left( #1 \right) }
         
         \def\X{{\rm X}}
         \def\L{{\rm L}}
         \def\R{{\rm R}}
         \def\M{{\rm M}}
         \def\H{{\rm H}}
         
         \def\T{{\cal T}}
         \def\Y{{\cal Y}}
         \def\Hc{{\cal H}}
         \def\kasr{{ 1 - q p^2 \over q - 1 }}
\begin{document}
\hfill
\vbox{
    \halign{#\hfil         \cr
            hep-th/9410135 \cr
            IPM-94-61      \cr
            SUTDP 94/73/9  \cr
            TUDP 94-1      \cr
            IASBS 94-5     \cr
           } 
      }  
\vskip 10 mm
\leftline{ \Large \bf
    $h$-Deformation as a Contraction of $q$-Deformation
         }
\vskip 10 mm
\leftline{ \bf
          A. Aghamohammadi ${}^{1,2,*}$,
          M. Khorrami ${}^{1,4,5}$, and
          A. Shariati ${}^{1,3}$
          }
\vskip 10 mm
{\it
\leftline{ $^1$ Institute for Studies in Theoretical Physics and Mathematics,
           P.O.Box  5746, Tehran 19395, Iran. }
  \leftline{ $^2$ Department of Physics, Alzahra University,
             Tehran 19834, Iran }
  \leftline{ $^3$ Department of Physics, Sharif University of Technology,
             P.O.Box 9161, Tehran 11365, Iran. }
  \leftline{ $^4$ Department of Physics, Tehran University,
             North-Kargar Ave. Tehran, Iran. }
  \leftline{ $^5$ Institute for Advanced Studies in Basic Sciences,
             P.O.Box 159, Gava Zang, Zangan 45195, Iran. }
  \leftline{ $^*$ E-Mail: mohamadi@vax.ipm.ac.ir}
  }
\begin{abstract}
We show that $h$-deformation can be obtained, by a singular limit of a
similarity transformation, from $q$-deformation; to be specefic, we obtain
$\GL_h(2)$, its differential structure, its inhomogenous extension, and
$\Uh{\sl(2)}$ from their $q$-deformed counterparts.
\end{abstract}
 { \large
\vskip 10 mm
The idea of using singular limits of transformations is not new.
Contraction of Lie groups, as first introduced by Inonu and Wigner \cite{A}
is such a process.
This contraction procedure has been successfully applied to quantum groups
to obtain deformations of inhomogeneous groups like \( {\rm E}_{q}(2) \) and
the  Poincar\'e group [2-4].

In this letter, we will show that $h$-deformation [5-13] can be obtained
from $q$-deformation by a singular limit of a similarity transformation.
Not only the Hopf algebras but also the whole
differential structure is obtained in this way.
Besides, this is also true for the inhomogeneous quantum group \( \IGL_h(2) \)
which can be obtained from \( \IGL_q(2) \).

Before describing the contraction procedure, it is worth mentioning why
\( \GL_q(2) \) and \( \GL_h(2) \) and their two-parametric generalizations are
imortant.
First, the only quantum groups which preserve nondegenerate bilinear forms
are \( \GL_{q \, p}(2) \) and \( \GL_{h \, h'}(2) \), \cite{E,L}.
These quantum groups act on the $q$-plane, with relation
\( x \, y = q \, y \, x \) and the $h$-plane, with relation
\( x \, y - y \, x = h \, y^2 \), respectively.
Second, it is shown that, up to isomorphisms, there exist only
two quantum deformations of \( \GL(2) \) which admit a central determinant
\cite{H}.
They are single parameter $q$- and $h$-deformations.
Finaly, \( 2 \times 2 \) quantum matrices admitting left and right quantum
spaces are classified \cite{M}.
They are two parametric $q$- and $h$-deformations.

To begin we define
\be
 \ba{ll}
   \M = \pmatrix{ a & b \cr c & d \cr }
   &
   \M' = \pmatrix{ a' & b' \cr c' & d' \cr }.
   \cr
 \ea
\ee
Throughout this letter we denote $q$-deformed objects by primed quantities.
Unprimed quantities represent similarity transformed objects, which, in a
certain limit, tend to $h$-deformed ones.
\( \M' \in \GL_{q \, p}(2) \), which means that the entries of
\( \M' \) fulfill the following commutation relations.
\be
 \label{C}
 \ba{lll}
   a'\, c'=qc'\, a' & b'\, d'=qd'\, b' &  [a',d']=qc'\, b'-q^{-1}b'\, c' \cr
   a'\, b'=q p^2 b'\, a' & c'\, d'=q p^2 d'\, c'  & c'\, b'= p^2 b'\, c' \cr
 \ea
\ee
\be
 D' = \det_q \M' :=a' \, d' - q \, c' b'
\ee
Here $[ \, , \, ]$ stands for the commutator.
\( \GL_{q \, p}(2) \) acts on the $q$-plane which is defined by
\be
 x' \, y' = q \, y' \, x'.
\ee
Now let us apply a change of basis in the coordinates of the $q$-plane
by use of the following matrix.
\be
 \label{A}
 g = \pmatrix{ 1 & { {h \over {q - 1} }} \cr 0 & 1 \cr }
 \qquad
\matrix{ x'= x + {h \over {q - 1 }} y \cr y' = y \cr }
\ee
A simple calculation shows that, the transformed generators $x$ and $y$
fulfill the following relation.
\be
 x \, y = q \, y \, x + h \, y^2.
\ee
The \( q \rightarrow 1 \) limit of this is exactly the commutation relation
that defines the $h$-plane:
\be
  x \, y = y \, x + h \, y^2.
\ee
The transformation on \( \GL_{q \, p}(2) \), corresponding to (\ref{A} ),
is the following similarity transformation.
\be
 \label{B}
 \ba{lll}
  \M' = g \, \M g^{-1} \cr
  a' = a + {h \over {q - 1}} c & &
  b' = b + { h \over {q - 1}} (d -a) - {h^2 \over {(q - 1)^2}} c \cr
  c' = c & &
  d' = d + { h \over {q - 1 } } c.
 \ea
\ee

It is clear that a change of basis in the quantum plane leads to the
similarity transformation $ \M = g^{-1} \M' g $ for the quantum group and
the following similarity transformation for the corresponding R-matrix.
\be
  \label{G}
  \R = ( g \otimes g)^{-1} \R' ( g \otimes g ).
\ee
We use the following R-matrix for the $q$-deformation.
\be
  \R' =
    \pmatrix{
               1  &  0     &  0           &  0  \cr
               0  &  qp^2  &  1-q^2 p^2   &  0  \cr
               0  &  0     &  q           &  0  \cr
               0  &  0     &  0           &  1  \cr
            } 
\ee
For $g$ given in (\ref{A}) we get
\be
  \R  =
    \pmatrix{
               1  &  h \kasr &  -hq \kasr   &  -h^2 \kasr \cr
               0  &  qp^2    &  1-q^2 p^2   &  -h q p^2   \cr
               0  &  0       &  q           &   h         \cr
               0  &  0       &  0           &   1         \cr
            } 
\ee
It is clear that $ \R \rightarrow \R_{h \, h'} $ as $ q \rightarrow 1 $
and $ p \rightarrow 1 $ provided that
\be
  \label{F}
{ { 1 - q p^2 } \over { 1 -q } } \rightarrow { h' \over h }.
\ee
\be
  \R_{h \, h'} =
    \pmatrix{
               1  &  -h'  &   h'   &   h h' \cr
               0  &   1   &   0    &  -h    \cr
               0  &   0   &   1    &   h    \cr
               0  &   0   &   0    &   1    \cr
            } 
\ee
The algebra of functions $\GL_{q \, p}(2)$ are obtained from the following
relation \be
  \R' \M'_1 \M'_2 = \M'_2 \M'_1 \R'
\ee
Appling transformations (\ref{B},\ref{G}) one obtains in the limit
\be
  \R \M_1 \M_2 = \M_2 \M_1 \R.
\ee
So, the entries of the transformed quantum matrix $\M$ fulfill the commutation
relations of the $\GL_{h \, h'}(2)$.
\be
 \ba{ll}
  [a,c]=hc^2    &
  [d,b]=h(D-d^2) \cr
  [a,d]=h d c-h' ac &
  [d,c]=h' c^2 \cr
  [b,c] =h' a c + h c d &
  [b,a] = h' (a^2 -D) \cr
 \ea
\ee
\be
 D = \det_h \M := a \, d - c \, b - h \, c \, d
\ee
One can obtain the above algebra by direct
substitution.
Note that although the transformations (\ref{A},\ref{B}) are ill behaved
as $q\rightarrow 1$ and $p\rightarrow 1$, the resulting commutation
relations are well defined in this limit. It is also important to note that
this process cannot be reversed: one can not use the inverse transformations
to obtain GL$_{q\, p}(2)$ from GL$_{h\, h'}(2)$.
The intuitive reason for this is that the coefficients of $xy$ and $yx$ in
the defining relation of the $h$-plane are equal; therefore, any change of
basis leads to the same coefficents and this is also true in any limit.
Because of this one-way nature of transformation, we call this process
a {\it contraction}. It is easily shown that the co-unity, antipode, and
co-product structures are also transformed to their $h$-deformed counterparts.
The inhomogeneous quantum group $\IGL_{q \, p}(2)$ has two extra
generators $u'$ and $v'$ which we arrange them in the matrix form:
\be
U' := \pmatrix{ u' \cr v' }.
\ee
The commutation relations
for these extra generators, which correspond to translations, are:
\be
 \ba{l}
    uv=qvu                \cr
    (av+uc)-q(cu+va)=0    \cr
    (bv+ud) -q(du+vb) =0  \cr
 \ea
\ee
Appling (\ref{A},\ref{B}) and $ U' = g U $ in the limit $ q \rightarrow 1 $
we get
\be
 \ba{l}
   uv-vu=hv^2                   \cr
   [b,v] + [u,d] = h \{d,v\}    \cr
   [a,v] + [u,c] = h \{ c,v \}  \cr
 \ea
\ee
where $ \{ \, , \, \} $ stands for the anticommutator.
These are the known commutation relations for $\IGL_{h \, h'}(2)$,
\cite{L}.

A quantum group's differential structure is completely determined by its
R-matrix \cite{N,O}. One therefore expects that by this similarity
transformation the differential structure of the $h$-deformation \cite{I} be
obtained from that of the $q$-deformation.
\be
 \ba{l}
   \M_2 d\M_1 = \R_{12} d\M_1 \M_2 \R_{21}  \cr
   d\M_2 d\M_1 + \R_{12} d\M_1 d\M_2 \R_{21} = 0
 \ea
\ee

Now, it is obvious that, defining $d\M := g^{-1} d \M' g$ and
using the above relations
the differential structure of $\GL_{h \, h'}(2)$ can be easily obtained
from the corresponding differential structure of $\GL_{q \, p}(2)$.

To proceed with the deformed universal enveloping algebras we use the well
known duality between $\Uq{\sl(2)}$ and $\SL_{q}(2)$. The algebra
$\Uq{\sl(2)}$ is generated by three generators $\X^{\pm}$ and $\H$.
They can be arranged in the following matrix form:
\be
 \ba{ll}
   \L'^{+} =
     \pmatrix{
      q^{-\half \H} &  (q - q^{-1}) \X^{+} \cr
      0                  & q^{\half \H}   \cr
              }
  &
   \L'^{-} =
     \pmatrix{
       q^{\half \H }     &   0                    \cr
       (q^{-1} - q ) \X^{-}   &   q^{-\half\H }  \cr
              }.
 \ea
\ee
$\Uq{\sl(2)}$ is generated by three generators $\X^{\pm}$ and $\H$. The
duality relations are:
\be
  \ba{ll}
    < \L'^{+}_{kl} , \M'_{ij} > = \R'^{+}_{ik \, jl}
    &
    < \L'^{-}_{kl} , \M'_{ij} > = \R'^{-}_{ik \, jl} .
    \cr
   \ea
\ee
Where
\be
  \ba{lll}
    \R'^{+} := \R'^{-1}
    &
    \R'^{-} := P \R' P
    &
    P_{ij \, mn} := \delta_{in} \delta_{jm}.
    \cr
  \ea
\ee
Note that we have to use these because we have begun with an upper triangular
$\R$-matrix. Now we use $g$ to get $\M$, $\R^{+}$ and $\R^{-}$.
\be
    \R^{\pm} = ( g \otimes g)^{-1} \R'^{\pm} ( g \otimes g )
\ee
Using these, it  can be shown that
\be
  \ba{l}
    \L^{+} = g^{-1} \L'^{+} g
    \cr \cr
    \L^{+} =
    \pmatrix{
      q^{-\half \H}
      &
      (q - q^{-1}) \X^{+} + \alpha ( q^{\half \H} - q^{-\half \H})
      \cr
      0
      &
      q^{\half \H}
      \cr
    }
   \cr \cr
    \L^{-} = g^{-1} \L'^{-} g
    \cr \cr
    \L^{-} =
    \pmatrix{
      q^{\half \H }+\alpha(q^{-1} -q) \X^{-}
      &
      -\alpha^2 (q^{-1}-q)\X^{-}+\alpha( q^{-\half \H} -q^{\half\H })
      \cr
      (q^{-1} - q) \X^{-}
      &
      q^{-\half \H }-\alpha(q^{-1} - q) \X^{-}
   }
   \cr
  \ea
\ee
where
\be
\alpha = { h \over q - 1 }
\ee
Now we introduce the following generators:
\be
  \ba{l}
    \T      := q^{ - {\H \over 2} } \cr
    \T^{-1} := q^{  {\H \over 2} } \cr
    \Hc     := \alpha  h^{-1} ( q^{{\H \over 2}}  - q^{-{\H \over 2}} )
               + h^{-1} ( q - q^{-1} ) \X^{+}  \cr
    \Y      :=  { 2\alpha \over ( q^{-1} - q ) } (q^{{\H \over 2}} -
               q^{-{\H \over 2}} ) - \X^{+} - {{4\alpha h} \over
               { (q+1) (q^{-1} - q)} } \X^{-} \cr
  \ea
\ee
It can be shown that these generators fulfill the following set of
commutation relations:
\be
  \ba{l}
    q \T \Hc - \Hc \T =  1 - \T^2                                      \cr
    q \Hc \T^{-1} - \T^{-1} \Hc =  \T^{-2} - 1                         \cr
    \Y \T - q^{-1} \T \Y = - { h \over q+1 } \{ \Hc , \T \}            \cr
    \Y \T^{-1} -  q \T^{-1} \Y =  { hq \over q+1 } \{ \Hc , \T^{-1} \} \cr
    [ \Hc , \Y ] = - { 1 \over q + 1 } \{ Y , \T^{-1} + \T \}.         \cr
  \ea
\ee
We see that the $ q \rightarrow 1$ limit of these equations are
the known relations for $\Uh{\sl(2)}$, \cite{G}. It is easy
to show that co-unity, co-product and antipode are also obtained in the
$ q \rightarrow 1$ limit from their corresponding $q$-deformed counterparts.

 } 

\end{document}